\documentclass[aps,prb,twocolumn,superscriptaddress,longbibliography]{revtex4-1}
\usepackage{epsfig,amsopn}
\usepackage{graphicx}
\usepackage{color}
\usepackage{amsmath,amssymb}
\usepackage{enumerate}
\newcommand\bea{\begin{eqnarray}}
\newcommand\eea{\end{eqnarray}}
\newcommand\beq{\begin{equation}}
\newcommand\eeq{\end{equation}}

\def\nn{\nonumber}
\def\f{\frac}

\def\si{\sigma}
\def\Do{\partial}

\def\ra{\rangle}
\def\ua{\uparrow}
\def\da{\downarrow}

\def\th{\theta}

\begin{document}
\title{Finite transverse conductance in topological insulators under an applied in-plane magnetic field}
\author{Dhavala Suri~~}
\affiliation{Tata Institute of Fundamental
Research, Hyderabad-500046, India}
\author{ Abhiram Soori~~}
\email{Corresponding author: abhirams@uohyd.ac.in}
\affiliation{School of Physics, University of Hyderabad, C. R. Rao Road, Gachibowli,
Hyderabad-500046, India}

\begin{abstract}
Recently, in topological insulators~(TIs) the phenomenon of planar Hall effect~(PHE) wherein a current driven in presence an in-plane magnetic field generates a transverse voltage has been experimentally witnessed. There have been a couple of theoretical explanations  of this phenomenon. We investigate this phenomenon based on scattering theory on a normal metal-TI-normal metal hybrid structure and calculate the conductances in longitudinal and transverse directions to the applied bias. The transverse conductance depends on the spatial location between the two NM-TI junctions where it is calculated. It is zero in the drain electrode when the chemical potentials of the top and the bottom TI surfaces ($\mu_t$ and $\mu_b$ respectively) are equal. The longitudinal conductance is $\pi$-periodic in $\phi$-the angle between the bias direction and the direction of the in-plane magnetic field. The transverse conductance is $\pi$-periodic in $\phi$ when $\mu_t=\mu_b$ whereas it is $2\pi$-periodic in $\phi$ when $\mu_t\neq\mu_b$. As a function of the magnetic field, the magnitude of transverse conductance increases initially and peaks. At higher magnetic fields, it  decays for angles $\phi$ closer to $0,\pi$ whereas oscillates for angles $\phi$ close to $\pi/2$. The conductances oscillate with the length of the TI region. A finite width of the system makes the transport separate into finitely many channels. The features of the conductances are similar to those in the limit of infinitely wide system except when the width is so small that only one channel participates in the transport. When only one channel participates in transport, the transverse conductance in the region $0<x<L$ is zero for $\mu_t=\mu_b$ and the transverse conductance in the region $x>L$ is zero even for the case $\mu_t\neq\mu_b$. We understand the features in the obtained results. 
\end{abstract}
\maketitle
\section{Introduction}
In the last few decades, novel materials such as topological insulators~(TIs) and Weyl semimetals which exhibit nontrivial electrical properties stemming from the topology of their bandstructures were predicted and realized~\cite{qi11,hasan10,yan17,armitage18}. Under an external magnetic field, a current driven results in development of a voltage transverse to the current in the plane of magnetic field and current, and this phenomenon is called planar Hall effect (PHE). PHE along with  negative longitudinal magnetoresistance has been seen as a direct signature of chiral anomaly in Weyl semimetals \cite{burkov17,nandy17,kumar18}. PHE has also been observed in TIs~\cite{taskin17,rakhmi18,he19,bharadwaj21} and its origin is ascribed to spin-flip scattering of surface electrons from impurities. Another explanation of PHE comes from the tilting of the Dirac cone that describes the surface states of the TIs~\cite{zheng20}. Also there has been an attempt at explaining PHE emanating from the bulk states of the TI~\cite{nandy18}. It is interesting to note that PHE in TIs  was predicted by considering scattering at junction of TIs with a ferromagnet in proximity to one part of the TI surface~\cite{scharf16}, without the need of  either the scattering from impurities or the titling of the Dirac cones due to magnetic field. But a TI has two surfaces- one on top and another at bottom, as a result, it is not clear whether the transverse deflections of the incident electrons will cancel from the two surfaces. Motivated by these developments, we examine transport in a system of in-plane magnetic field applied to top and bottom surfaces of a TI connected to  two-dimensional normal metal~(NM) leads on either sides. We follow Landuer-B\"uttiker approach~\cite{landauer1957r,buttiker1985m,datta1995} and calculate currents in transverse and longitudinal directions in response to a bias applied in the longitudinal direction. This is in contrast to the experiments where a current is driven in longitudinal direction and voltages developed in transverse and longitudinal directions are measured in Hall bar geometry. Also, we study the effect of unequal chemical potentials on the top and the bottom surfaces of TI which can be achieved in experiments by applying different gate voltages to the two surfaces. Finally, we study the case of finite width of the sample. 

\begin{figure}[htb]
 \includegraphics[width=7cm]{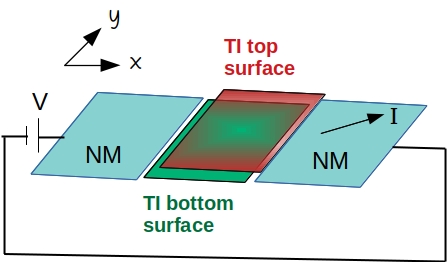}
 \caption{Schematic diagram of the setup: the topological insulator~(TI) is connected to normal metal~(NM) leads on either sides.   The two NM's and the TI are taken to be infinitely long along $y$. Both the  NM leads are semi-infinite along $x$. A voltage $V$ applied from left NM to the right NM results in a current $I$. Planar Hall effect is when the current $I$ has a nonzero component along $y$ direction. Such a transverse current could be either in the TI region or in the right NM region.}\label{fig-schem}
\end{figure}

The paper is organized as follows. In sec.~\ref{sec:calc}, the system under consideration and details of the calculation comprising of the Hamiltonian, the boundary conditions and the formulae for the longitudinal and the transverse conductances  are discussed. In sec.~\ref{sec:result}, the results are presented and analyzed. In sec.~\ref{sec:conclude}, we discuss the implications of our results and conclude. 

\section{Details of calculation}~\label{sec:calc}

The setup under study is a NM-TI-NM junction, with the TI in the middle having a top surface and a bottom surface as shown in the Fig.~\ref{fig-schem}. We shall take both the NMs and the TI to be of length $L_y$ along $y$. The NM lead on the left extends all the way from $x=-\infty$ to $x=0$ and makes a junction with both the surfaces of TI along  $x=0$. TI extends from $x=0$ to $x=L$ and makes a junction with the NM on the right along the line $x=L$. From now on, we shall denote the coordinates of the top~(bottom) surface of the TI with a subscript $t$~($b$). The in-plane magnetic field applied is present only in the TI region.  The NM lead on the right extends from $x=L$ to $x=\infty$. The Hamiltonian describing the system being investigated is 
\begin{widetext}
\bea H &=& \Big[-\f{\hbar^2}{2m}\Big(\f{\Do^2}{\Do x^2}+\f{\Do^2}{\Do y^2}\Big)-\mu_N\Big]\si_0, {\rm ~~for~}x<0{~\rm and~}x>L, \nn \\ &=&  i\hbar v_F\Big(\si_y\f{\Do}{\Do x_t}-\si_x\f{\Do}{\Do y_t}\Big)+(b_x\cdot\si_x+b_y\cdot\si_y)-\mu_t\si_0, {\rm ~~for~}0<x_t<L,\nn \\ &=&  -i\hbar v_F\Big(\si_y\f{\Do}{\Do x_b}-\si_x\f{\Do}{\Do y_b}\Big)+(b_x\cdot\si_x+b_y\cdot\si_y)-\mu_b\si_0, {\rm ~~for~}0<x_b<L. \label{eq:ham} \eea
\end{widetext}
Here, $\mu_N$ is the chemical potential of the NM leads,  $\mu_{t/b}$ is the chemical potential on the top/bottom surface of the TI which can be controlled by  applied gate voltages, 
$(b_x,b_y)=b(\cos\phi,\sin\phi)$ where $\phi$ is the angle the in-plane magnetic field makes with $x$-axis (we refer to the Zeeman energy $b$ as magnetic field), $\si_0$ is identity matrix and $\si_{x,y}$ are Pauli spin matrices. The Hamiltonians for the top and the bottom surfaces have a relative minus sign for the following reason. The two surfaces are part of the same TI and are connected at the boundaries. If $L_y$ is the length of the TI in $y$-direction, we can think of the top and the bottom surfaces of TI as a single TI surface described by the Hamiltonian $i\hbar v_F(\si_y\Do_x-\si_x\Do_y)$ along with the periodic boundary conditions: $x=0\equiv 2L$ and  $y=0\equiv 2L_y$. The coordinates of the top and the bottom surfaces are $x_t=x$ in the range $0<x<L$, $x_b=2L-x$ in the range $L<x<2L$, $y_t=y$ in the range $0<y<L_y$ and $y_b=2L_y-y$ in the range $L_y<y<2L_y$, which imply $\Do_{x_b}=-\Do_x$ and $\Do_{y_b}=-\Do_y$ leading to the relative minus sign. This can also be shown starting from the bulk four band Hamiltonian~\cite{udupa18}. Though the bulk TI Hamiltonian has four bands, two coming from spin and another two coming from  bipartite nature of the underlying lattice, the magnetic field couples only to the spin through Zeeman coupling resulting in the term $(b_x\si_x+b_y\si_y)$.  We have chosen the gauge for the vector potential so that it is zero in $(x,y)$ plane: $\vec A =(0,0,b_xy-b_yx)$.
The in-plane magnetic field shifts the Dirac point of the top~(bottom) surface to $\vec k =\pm (b_y,-b_x)/\hbar v_F$ respectively.  The dispersion relations for the top and the bottom TI surfaces are respectively 
\bea E&=&-\mu_t\pm\sqrt{(\hbar v_Fk_x-b_y)^2+(\hbar v_Fk_y+b_x)^2}, ~~~~\label{eq:disp-t} \\E&=&-\mu_b\pm\sqrt{(\hbar v_Fk_x+b_y)^2+(\hbar v_Fk_y-b_x)^2} ~.~~~\label{eq:disp-b} \eea

To solve the scattering problem, boundary conditions need to be specified at $x=0$ and $x=L$. Boundary conditions at NM-TI junctions have been discussed in literature~\cite{modak12,soori13,soori20}.  The probability current operators for the top and bottom surfaces can be shown to be $\hat \vec j_t=(-v_F\si_y,v_F\si_x)$ and  $\hat \vec j_b=(v_F\si_y,-v_F\si_x)$ respectively. So, the conservation of current along $x$-direction between NM and TI surfaces reads 
\bea \f{\hbar~ {\rm Im}[\psi^{\dagger}_N\Do_x\psi_N]_{x=x_0}}{mv_F} &=& -\psi^{\dagger}_t\si_y\psi_t|_{x_t=x_0} + \psi^{\dagger}_b\si_y\psi_b|_{x_b=x_0},~  \nn \\ &&\label{eq:current-cons}\eea
at both the junctions located at $x_0=0,~L$, where $\psi_N$ is the wavefunction on the NM side and  $\psi_{t/b}$ is the wavefunction on the top/bottom surface of the TI. The most general boundary condition satisfying the current conservation eq.~\eqref{eq:current-cons} is
\bea \psi_N &=& c[M(-\chi_t)\psi_t+M(\chi_b)\psi_b], \nn \\ \f{\hbar}{mv_F}\Do_x\psi_N-\chi_{N}\psi_N &=& \f{i}{c}\si_y\cdot\Big[-M(-\chi_t)\psi_t + M(\chi_b)\psi_b\Big], \nn \\ && \label{eq:bc} \eea
where all the wavefunctions and $\Do_x\psi_N$ are evaluated at the junction at $x=0$. Here, $M(\chi)={\exp}[i\chi\si_y]$. We shall soon see that the dimensionless parameters $\chi_N$, $\chi_t$ and $\chi_b$ quantify the strengths of the delta-function barriers infinitesimally close to the junction from the NM-, top~TI- and bottom~TI- sides respectively~\cite{modak12,sen12,*sen12err}. The boundary conditions at the junction located at $x=L$ is same as eq.~\eqref{eq:bc}, except that the dimensionless parameters $\chi_N$, $\chi_t$ and $\chi_b$ acquire opposite signs. A delta function barrier on the NM side of a junction results in a wavefunction which continuous at the location of the barrier, accompanied by a discontinuity in $\Do_x\psi$ proportional to the strength of the barrier multiplied by $\psi$. Hence, $\chi_N$ is the strength of the barrier  on the NM side made dimensionless. On a TI surface described by the Hamiltonian $H_{TI}=i\hbar v_F(\si_y\Do_x-\si_x\Do_y)+ V_0 \Delta_l(x-x_0)$ (where $\Delta_l(x-x_0)$ is $1$ for $x_0<x<x_0+l$ and $0$ elsewhere), the wavefunction in the region $x_0<x<x_0+l$ obeys $i\hbar v_F\si_y\Do_x\psi=-V_0\psi$ for large $V_0$ and has a solution of the form $\psi(x)=\exp[iV_0 x\si_y/(\hbar v_F)]\psi(x_0)$. The delta function limit is $l\to0$ and $V_0\to\infty$ so that $V_0l$ is a finite constant. In this limit, $\psi(x_0^+)=\exp(i\chi_0\si_y)\psi(x_0^-)$, where $\chi_0=V_0l/(\hbar v_F)$. So, the wavefunction of the top/bottom TI surface across a delta function  barrier is related by $\psi(x_0^+)=\exp(\pm i\chi_0\si_y)\psi(x_0^-)$. This justifies the introduction of parameters $\chi_t$ and $\chi_b$ in the boundary conditions eq.~\eqref{eq:bc}.  We shall set all the dimensionless barrier strengths close to the junction to zero at both the junctions to allow  maximal transmission. We shall set $c=(mv_F/\sqrt{2m\mu_N})^{1/2}$ so that transmission of normally incident electron at the junction is perfect at zero energy in absence of a magnetic field~\cite{soori20}.  

Due to translational invariance of the system in $y$-direction, the momentum $\hbar k_y$ along $y$ can be taken to be equal in all the four regions. The component of the current along $x$ is conserved and is same anywhere. But along $y$, $k_y$ is same in all the regions and the component of current along $y$ need not be same at all $x$. 
\subsection{Limit as $L_y\to\infty$}
The wavefunction of a spin-$\si$ electron incident from the left NM with energy $E$, making an angle $\th$ with $x$-axis  has the following form in different regions (except for a multiplicative factor of $e^{ik_yy}$):
\bea \psi_N(x) &=& (e^{ik_xx}+r_{\si,\si}e^{-ik_xx})|\si\ra+r_{\overline{\si},\si}e^{-ik_xx}|\overline{\si}\ra,\nn \\&&~~~~~~~~~~~~~~~~~~ {~\rm for~}x<0, \nn \\ \psi_p(x_p) &=& s_{\si,p,+}e^{ik_{x,p,+}x_p}|k_{p,+}\ra + s_{\si,p,-}e^{ik_{x,p,-}x_p}|k_{p,-}\ra, \nn \\ &&~~~~~~~~~~~~~~~~~~{~\rm for~}0<x_p<L,~{\rm and~} p=t,b, \nn \\ \psi_{N}(x) &=& t_{\ua,\si}e^{ik_xx}|\ua\ra + t_{\da,\si}e^{ik_xx}|\da\ra, ~{\rm for~}x>L,~\label{eq:psi} \eea
where $\si=\ua,\da$, $\overline{\si}$ is the spin opposite to $\si$, $|\ua\ra=[1,0]^T$, $|\da\ra=[0,1]^T$, $k_x=\sqrt{2m(\mu_N+E)}\cos{\th}/\hbar$, $k_y=\sqrt{2m(\mu_N+E)}\sin{\th}/\hbar$, $k_{x,p,s}$'s for $s=+,-$ correspond to the two roots for $x$-wavenumber obtained from the dispersion in the $p$-TI surface  ($p=t,b$ stand for top, bottom surfaces) as a function of $E$ and $k_y$, $|k_{p,s}\ra$ is the spinor on $p$-TI surface for electron with wavenumber $(k_{x,p,s},k_y)$ which can be found from the Hamiltonian for the TI and the coefficients $r_{\si',\si}$, $s_{\si,p,s}$, $t_{\si',\si}$ are to be determined by matching the boundary conditions in eq.~\eqref{eq:bc} at $x_0=0,L$. 

  If $\psi_{p,\si}(x)$ is the wavefunction due to an  $\si$-spin electron incident at an angle $\th$ at energy $E$ on $p$-TI surface at $x_p=x$ in the range $0\le x_p\le L$, the current along $y$ at the location $x$ from this wavefunction will be $I_{\si,y}(E,\th,x)=ev_F\sum_{p=t,b}\psi_{p,\si}(x)^{\dagger}\si_p\si_x\psi_{p,\si}(x)$, where $e$ is electron charge,  $\si_t=1$ and $\si_b=-1$. If $[I_x,I_y(x)]$ is the current flowing at $x$ in response to a voltage bias $V$ in the bias window $(0,eV)$, the longitudinal and transverse differential conductances are defined as $G_{xx}=dI_x/dV$ and $G_{yx}(x)=dI_y(x)/dV$ respectively. These are given by the expressions 
\bea G_{xx} &=& \f{\sqrt{2m(\mu_N+eV)}}{mv_F} G_0\sum_{\si,\si'=\ua,\da}\int_{-\pi/2}^{\pi/2}d\th\cos{\th} |t_{\si',\si}|^2, \nn \\ G_{yx}(x) &=& \f{G_0}{ev_F}\sum_{\si=\ua,\da}\int_{-\pi/2}^{\pi/2}I_{\si,y}(eV,\th,x) d\th,~{\rm for ~}0<x<L, \nn \\ && \label{eq:Gxx-Gyx} \eea
where $G_0=(e^2/h)\cdot(mv_FL_y/h)$ and $L_y$ is the length of the system in $y$-direction. The current deflected in the transverse direction in the right NM is same at all locations $x>L$ and the transverse differential conductance due to this current is given by 
\bea && G_{yx}(x>L) \nn \\  &=& \f{\sqrt{2m(\mu_N+eV)}}{mv_F} G_0\sum_{\si,\si'=\ua,\da}\int_{-\pi/2}^{\pi/2}d\th\sin{\th} |t_{\si',\si}|^2~.  \label{eq:Gyxn} \eea
\subsection{Finite $L_y$}
For a finite $L_y$, we take the same Hamiltonian as in eq.~\eqref{eq:ham}, make the length along $y$-direction in all the regions $L_y$ finite  and apply periodic boundary conditions along $y$. This makes $k_y$ take values: $k_y=n2\pi/L_y$, for integer $n$. The scattering problem becomes one-dimensional and separated in channels described by $n$. At a given energy $E$, there are a finite number of channels participating in the transport given by $(2N+1)$, where $N=[\sqrt{2m(E+\mu_N)}L_y/h]$, $[x]$ being the largest integer less than $x$. For a given $k_y=n2\pi/L_y$ at energy $E$, $k_x=\sqrt{2m(E+\mu_N)/\hbar^2-k_y^2}$ and the wavefunction is given by eq.~\eqref{eq:psi} except that the wavefunction and the scattering coefficients carry an additional channel index $n$. Transverse current $I_{\si,y,n}$ carried by the channel indexed by $n$ due to an incident spin $\si$ electron is $I_{\si,y,n}=ev_F(\psi_{t,n}^{\dagger}\si_x\psi_{t,n}-\psi_{b,n}^{\dagger}\si_x\psi_{b,n})$. The longitudinal and the transverse conductances are given by 
\bea G_{xx} &=& \f{e^2}{h}\sum_{n=-N}^N\sum_{\si,\si'}|t_{\si',\si,n}|^2 \nn \\ G_{yx}(x)&=& G_0\f{2\pi}{L_y}\sum_{n=-N}^N\sum_{\si}\f{I_{\si,y,n}}{ev_F}, ~~{\rm for ~}0<x<L, \nn \\ G_{yx}(x>L)&=& \f{e^2}{h}\sum_{n=-N}^N\sum_{\si,\si'}\f{k_y}{k_x}|t_{\si',\si,n}|^2~. \label{eq:G-finite} \eea

\section{Results and Analysis}~\label{sec:result}
To obtain numerical results, we shall fix $\mu_N$ and $v_F$, and choose other parameters as combinations of these parameters.  The mass $m$ decides the size of the Fermi wavenumber. We choose $m=0.025\mu_N/v_F^2$ so that the wavenumbers on NM and TI at energy $-0.2\mu_N$ are equal when $\mu_t=\mu_b=0$ and $b=0$. The length of the TI is chosen to be $L=5\hbar v_F/\mu_N$. These are the values of the parameters unless otherwise stated. First we will discuss the results for the case $\lim{L_y\to\infty}$ and deliberate  upon the effect of finite $L_y$ at the end. 
\subsection{$\mu_t=\mu_b$}
\begin{figure*}[htb]
  \includegraphics[width=5.6cm]{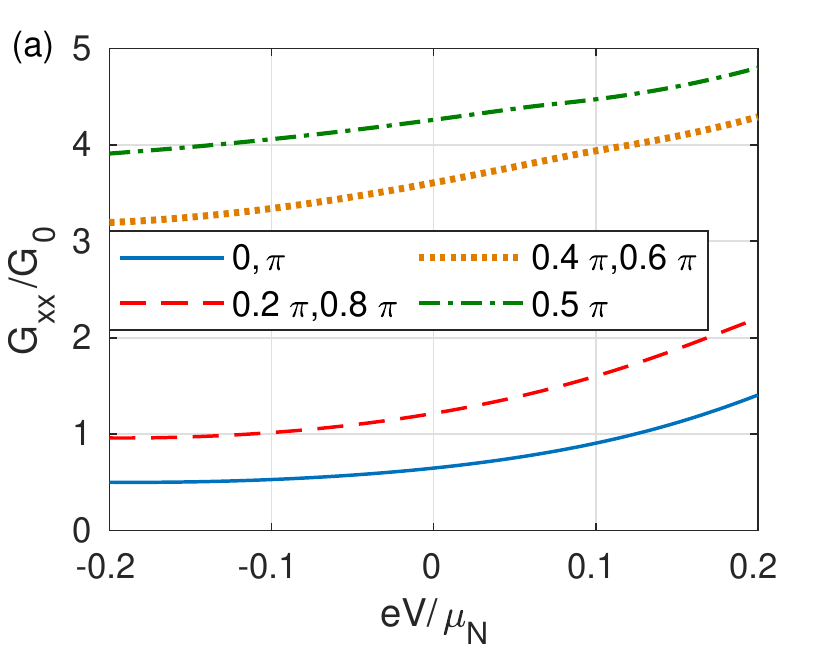}
  \includegraphics[width=5.6cm]{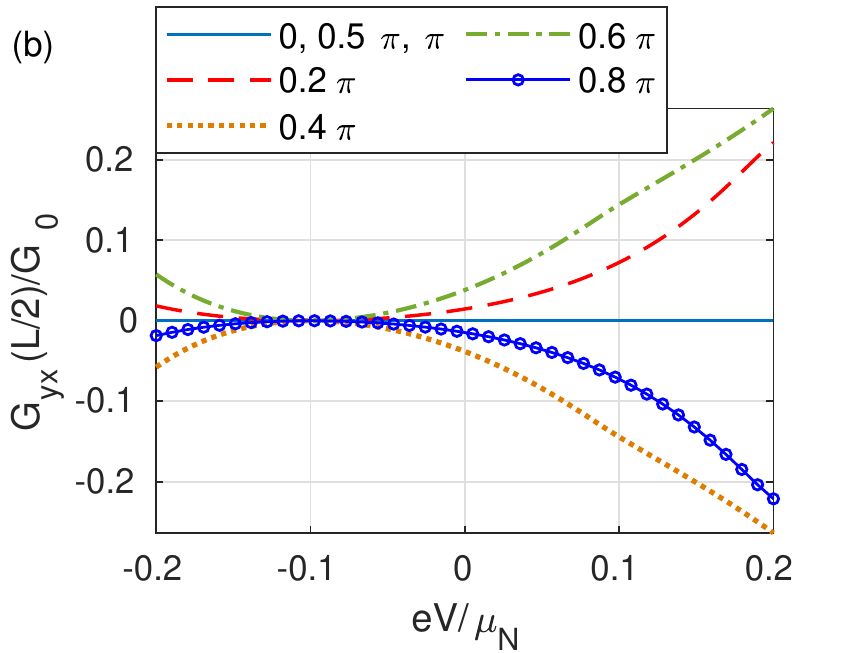}
  \includegraphics[width=5.6cm]{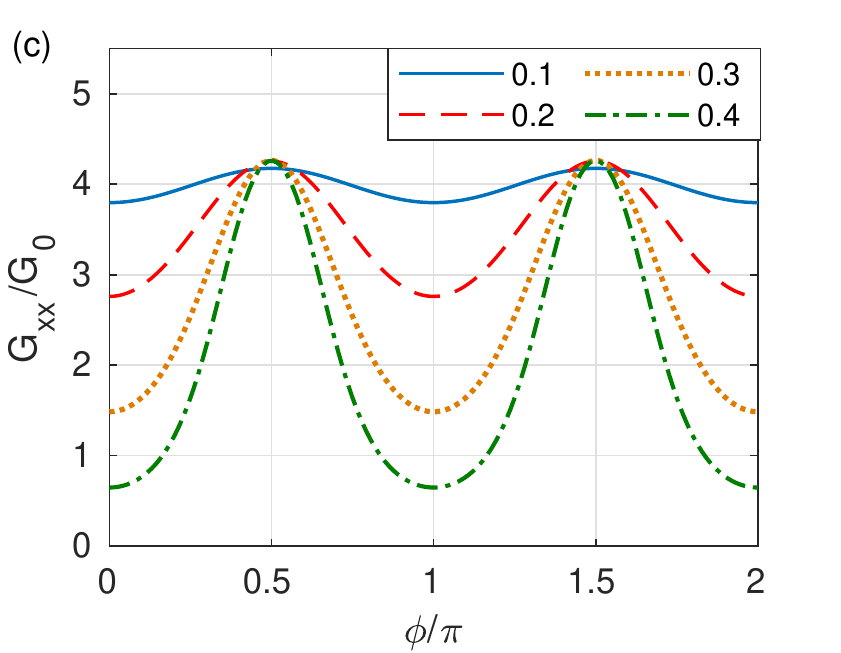}  
  \includegraphics[width=5.6cm]{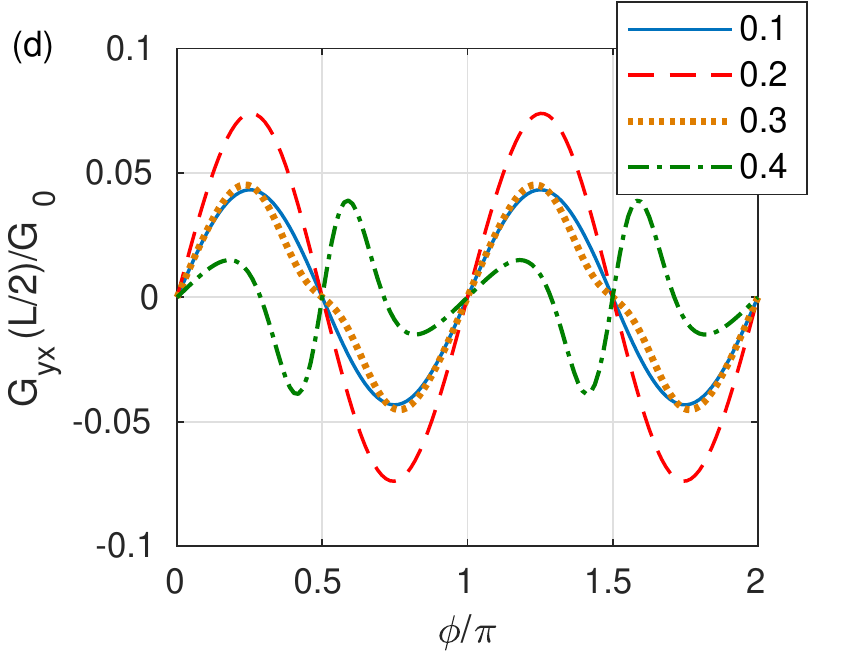}
  \includegraphics[width=5.6cm]{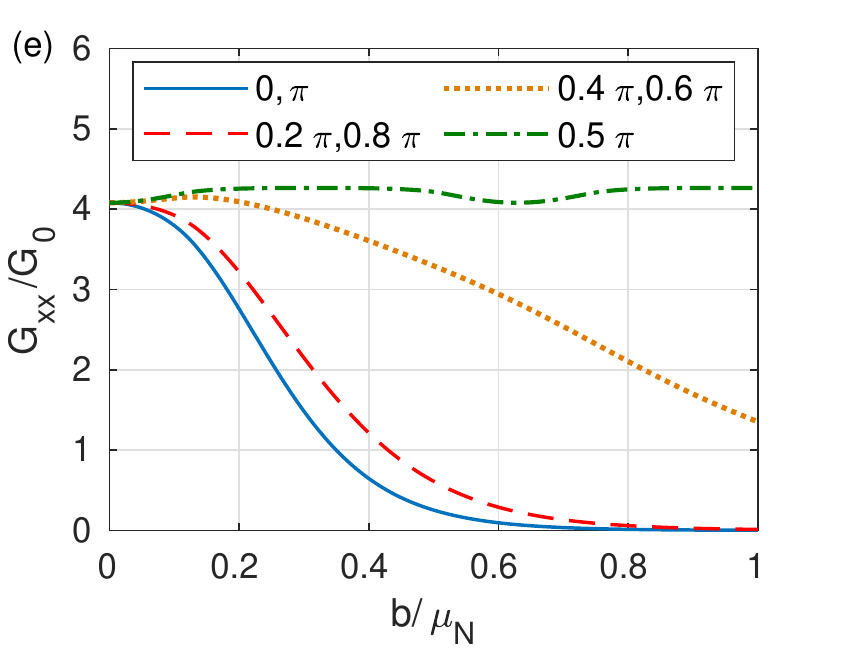}  
  \includegraphics[width=5.6cm]{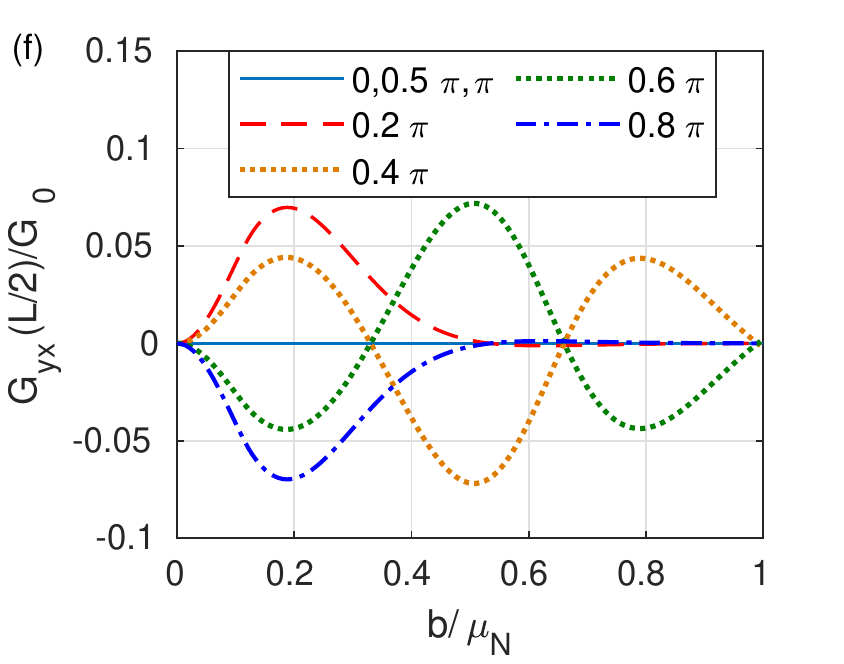}
  \caption{(a) $G_{xx}$ and (b) $G_{yx}$ as  functions of bias for different angles $\phi$ made by the in-plane magnetic field with $x$-axis with $b=0.4\mu_N$. The values shown in the legend for (a) and (b) are the respective values of $\phi$ for which $G_{xx}$ and $G_{yx}(L/2)$ are plotted.  (c)~$G_{xx}$ and (d)~$G_{yx}$ at zero bias as functions of $\phi$ at different values of magnetic field $b$ mentioned within the plot legend.  (e)~$G_{xx}$ and (f)~$G_{yx}$ at zero bias as functions of magnetic field $b$ for different $\phi$ specified in the plot legends. Parameters: $L=5\hbar v_F/\mu_N$,  $\mu_t=\mu_b=0.1\mu_N$ and $m=0.025\mu_N/v_F^2$. }\label{fig:gvsVphib}
 \end{figure*}
 
First,  we set $\mu_t=\mu_b=0.1\mu_N$ and study the dependence of $G_{xx}$ and $G_{yx}(L/2)$ on the bias at different angles $\phi$ when the magnitude of the magnetic field is fixed at $b=0.4\mu_N$ in Fig~\ref{fig:gvsVphib}(a) and Fig.~\ref{fig:gvsVphib}(b) respectively. In Fig.~\ref{fig:gvsVphib}(c) and Fig.~\ref{fig:gvsVphib}(d) we show the dependence of the longitudinal and the transverse conductances respectively at zero bias on $\phi$.    The slow increase in $G_{xx}$ with bias is due to increase in density of states of incident electrons with bias.
For an angle $\phi$ between $x$-axis and the magnetic field, the Dirac cones on the TI surfaces are displaced in $y$-direction by an amount $|b\cos{\phi}/(\hbar v_F)|$ thereby making the wavenumbers $k_{x,p,s}$'s in the TI region complex (when $\cos{\phi}\neq 0$) for a range of angle of incidence $\th$. 
This reduces the transmission probabilities $|t_{\si,\si'}|^2$ for larger values of $|\cos{\phi}|$ and for larger values of $|b|$ which agrees with the observed features of $G_{xx}$ in Fig.~\ref{fig:gvsVphib}(a,c). In Fig.~\ref{fig:gvsVphib}(b,d), we find that $G_{yx}(L/2)$ is exactly zero at $\phi=0,\pm\pi/2,\pi$. 
When $\phi=\pm\pi/2$, the Dirac cone is shifted along $x$-direction and the system is symmetric under $y\to-y$ thereby giving zero total current along $y$. When $\phi=0,\pi$, the Dirac cones in the top and bottom surfaces are displaced exactly along $\pm y$ directions, and the currents deflected along $y$ in the top and the bottom surfaces are equal in magnitude and opposite in sign thus giving zero $G_{yx}$. 
Now, we address the question why there is a nonzero $G_{yx}(L/2)$ for a nonzero $b$ in a direction $\phi$ other than $0,\pm\pi/2,\pi$. Under a magnetic field $(b_x,b_y)$, the Dirac points of the top and the bottom surfaces are shifted to $\pm(b_y,-b_x)/(\hbar v_F)$. The currents in $y$-direction carried by the electrons incident at angles $\th$ and $-\th$ on one surface do not cancel due to a finite shift of the Dirac cone in $y$-direction. At the same time, the net current in $y$-direction carried by the top surface and the bottom surface do not cancel despite the opposite shifts of the two Dirac cones because the wavenumbers in $x$-direction of the corresponding surfaces $k_{x,t,s}$ and $k_{x,b,s}$ are different. From Fig.~\ref{fig:gvsVphib}(b), it can be seen that at $eV=-\mu_t=-\mu_b$, the transverse conductance is exactly zero implying that the net current in the transverse direction carried by the evanescent waves in the TI region is zero. The transverse conductance $G_{yx}(x)$ is $\pi$-periodic in $\phi$ and  $G_{yx}(x>L)$ is exactly zero for the case $\mu_t=\mu_b$.

 Turning to the dependence of the two conductances on $b$,  in Fig.~\ref{fig:gvsVphib}(e), we find monotonic dependence of $G_{xx}$ on $|b|$ for $|\cos{\phi}|$ close to $1$ and oscillatory dependence of $G_{xx}$ on $|b|$ for $|\cos{\phi}|$ small compared to $1$. 
This is because, the displacement of TI Dirac cones on in $y$ direction is by an amount proportional to $\cos{\phi}$. Nearly normal incidences with $\th$ close to zero contribute the most to $G_{xx}$.   When $|\cos{\phi}|$ is large, for angles of incidences $\th$ close to zero, the transport in TI region is diffusive characterized by a complex $k_{x,p,s}$  whose imaginary part grows in magnitude with $b$. When $|\cos{\phi}|$ is small compared to $1$, the displacement of the TI Dirac cones along $y$-direction is minimal. Nearly normal incidences from NM will find a real $k_{x,p,s}$ in the TI and the transport is ballistic except for scatterings at the interfaces which leads to interference  between the forward moving and the backward moving waves. This is the reason for oscillatory behavior of $G_{xx}$ with $b$.  Under the transformation $\phi\to\phi+\pi$, the transmission probabilities $|t_{\si,\si'}|^2$ for angles of incidence $\th$ and $-\th$ get interchanged thereby making $G_{xx}$ even in $b$. In Fig.~\ref{fig:gvsVphib}(f), we plot $G_{yx}(L/2)$ versus $b$.  The nonzero values of the  transverse conductance $G_{yx}$ at certain values of $\phi$ increases in magnitude with $|b|$ for small $|b|$ since increasing value of $|b|$ gives scope for higher asymmetry between scatterings at angles of incidence $\th$ and $-\th$. But beyond a value of $|b|$, the displacement of the  Dirac cone in $y$-direction causes the wavefunction to decay into the TI (which is particularly the case for $|\cos{\phi}|$ close to $1$), resulting in decrease in magnitude of $G_{yx}$ with $|b|$. For values of $\phi$ such that $|\cos{\phi}|$ is small compared to $1$, the scattering from angles of incidence away from normal incidence centered around $\pm\th_b$ which depend on $|b|$ contribute dominantly to $G_{yx}$. The Fabry-P\'erot type interference~\cite{soori12} of these modes results in oscillations in $G_{yx}$ with $|b|$. Under $\phi\to\phi+\pi$, the displacement of each of the Dirac cones is opposite to that before the transformation. This hints at the reversal of sign of $G_{yx}$ upon $b\to-b$. But, since $b_y\to -b_y$ the displacement of each of the Dirac cones along $x$ is opposite to that before the transformation making the surface dominantly contributing to $G_{yx}$ switch under the transformation $\phi\to \phi+\pi$. Hence the displacement of the Dirac cone along $y$-direction for the surface dominantly contributing to $G_{yx}$ is shifted in the same direction for both choices of magnetic field directions $\phi$ and $\phi+\pi$,  making  $G_{yx}$ $\pi$-periodic. Further, the transmission probability at angle of incidence $\th$ for $\phi$ is equal to the transmission probability at angle $-\th$ for $\phi+\pi$ since under these transformations, the top and bottom surface Hamiltonians and the respective $k_{x,p,s}$'s  get interchanged [see eq.s~\eqref{eq:disp-b}\&\eqref{eq:disp-t}] leaving the transport problem along $x$ unchanged.  Hence, it can be seen from eq.~\eqref{eq:Gyxn} that transverse conductance in the NM region $G_{yx}(x>L)$ reverses sign under $\phi\to\phi+\pi$. This combined with $\pi$-periodicity of $G_{yx}$ implies $G_{yx}(x>L)$ is zero when the two chemical potentials are equal. 
\begin{figure}[htb]
\includegraphics[width=6cm]{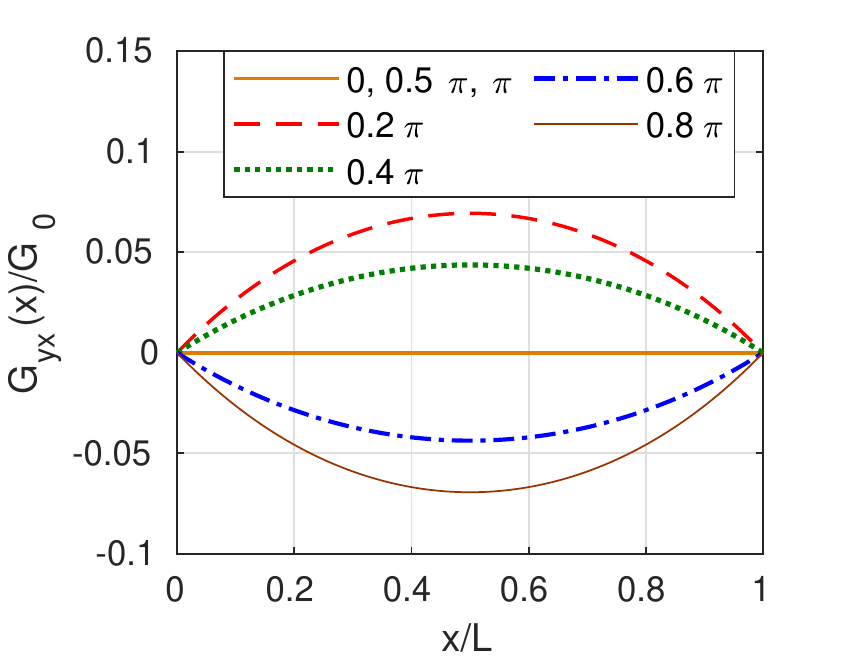}
 \caption{Dependence of the zero bias transverse differential conductance on the location in the TI region for different angles $\phi$ mentioned in the legend with the choice of parameters: $L=5\hbar v_F/\mu_N$,  $b=0.2\mu_N$, $\mu_t=\mu_b=0.1\mu_N$ and $m=0.025\mu_N/v_F^2$.}\label{fig:gyxvsx}
\end{figure}
To study the dependence of the transverse conductance on the location, we plot $G_{yx}(x)$ versus $x$ in Fig.~\ref{fig:gyxvsx}.  We find that the magnitude of the transverse conductance is peaked at $x=L/2$ for this choice of parameters. 

\subsection{$\mu_t\neq\mu_b$}
\begin{figure}[htb]
 \includegraphics[width=4.2cm]{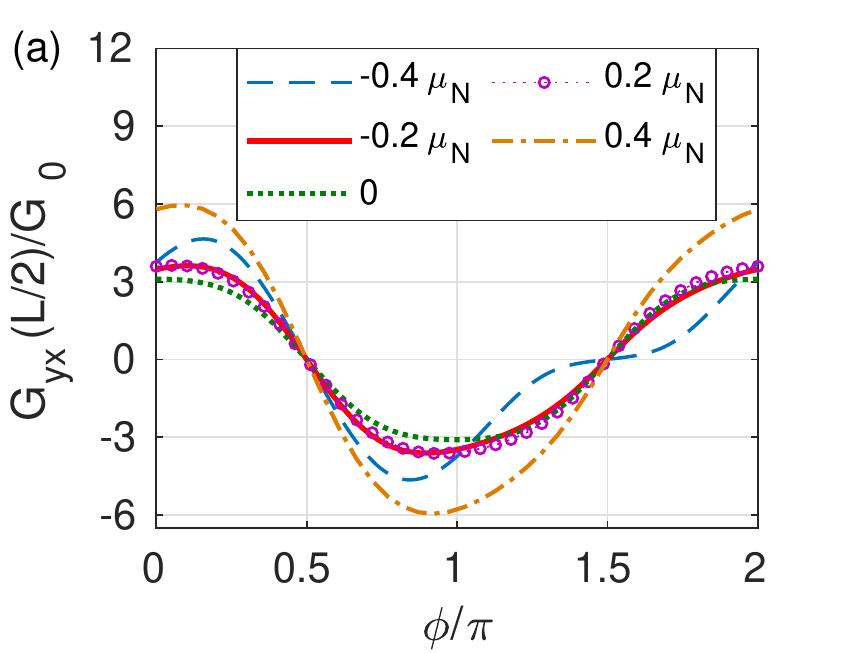}
\includegraphics[width=4.2cm]{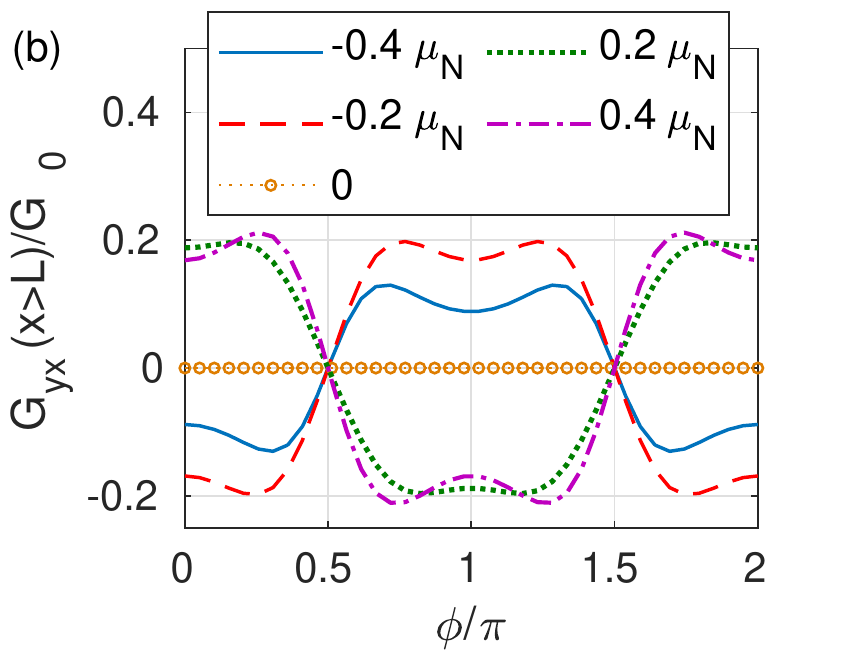}
\caption{Transverse differential conductance $G_{yx}(x)$ at (a) $x=L/2$ and at (b) $x>L$ as functions of $\phi$ for different values of bias mentioned in the legend. $\mu_t=-\mu_b=0.1\mu_N$, $b=0.2\mu_N$, $L=5\hbar v_F/\mu_N$,  and $m=0.025\mu_N/v_F^2$. }\label{fig:gyxvsphi}
 \end{figure}

 To study the conductances in this case, we choose the same set of parameters as in the Fig.~\ref{fig:gvsVphib} except when mentioned otherwise. We choose $\mu_t=-\mu_b=0.1\mu_N$.  The longitudinal differential conductance shows characteristics very similar to the ones in Fig.~\ref{fig:gvsVphib}(c) except for a change in the numerical value. Even in the case $\mu_t\neq\mu_b$, the longitudinal conductance is still $\pi$-periodic in $\phi$. The $\pi$-periodic behavior of longitudinal conductance can be understood as follows. Under the transformation $\phi\to(\phi+\pi)$, $(b_x,b_y)\to-(b_x,b_y)$ and the Dirac cones of the TI-surfaces get displaced exactly by the same magnitude but in opposite direction away from the origin in the $(k_x,k_y)$ plane. The transverse shift in opposite direction does not alter the longitudinal conductance. Furthermore, the longitudinal shift of Dirac cones in the opposite direction to the same extent does not alter the longitudinal conductance because, this is minus of  the longitudinal conductance when the same bias is applied in the opposite direction before reversing the magnetic field and the net longitudinal conductance at zero applied bias is exactly zero. In Fig.\ref{fig:gyxvsphi}, we plot the transverse differential conductances at $x=L/2$ and at $x>L$ versus $\phi$. 
It is interesting to see that $G_{yx}(x)$ is $2\pi$-periodic for $\mu_t=-\mu_b$. Also, $G_{yx}(x>L)$ is nonzero generically except at zero bias. Also, interestingly both $G_{yx}(L/2)$ and $G_{yx}(x>L)$ are nonzero at $\phi=0$ for this case. This  is because of the  displacement of the two Dirac cones in $\pm y$-direction  equally but in opposite directions does not lead to cancellation of transverse currents at nonzero bias due to broken perfect antisymmetry of the top-bottom surface dispersions under $y\to-y$. For a fixed bias, the values of $G_{yx}(x)$ for $\phi$ and $\pi-\phi$ are equal in magnitude and opposite in sign since the transverse shift of the Dirac cones is exactly opposite for $\phi\to(\pi-\phi)$. 
 The transverse conductance is $\pi$-periodic only when the chemical potentials of the top and bottom surfaces are the same since under $\phi\to\phi+\pi$, the Dirac cones of the top and the bottom surfaces get interchanged, whereas when  $\mu_t\neq \mu_b$, under $\phi\to\phi+\pi$  the top and the bottom Dirac cones do not get interchanged.

\begin{figure}[htb]
 \includegraphics[width=4.2cm]{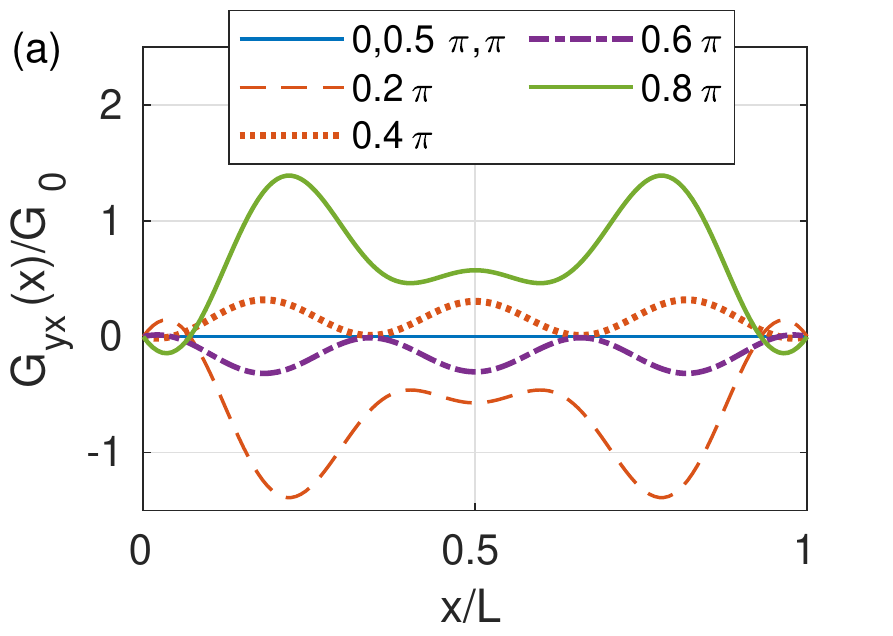}
  \includegraphics[width=4.2cm]{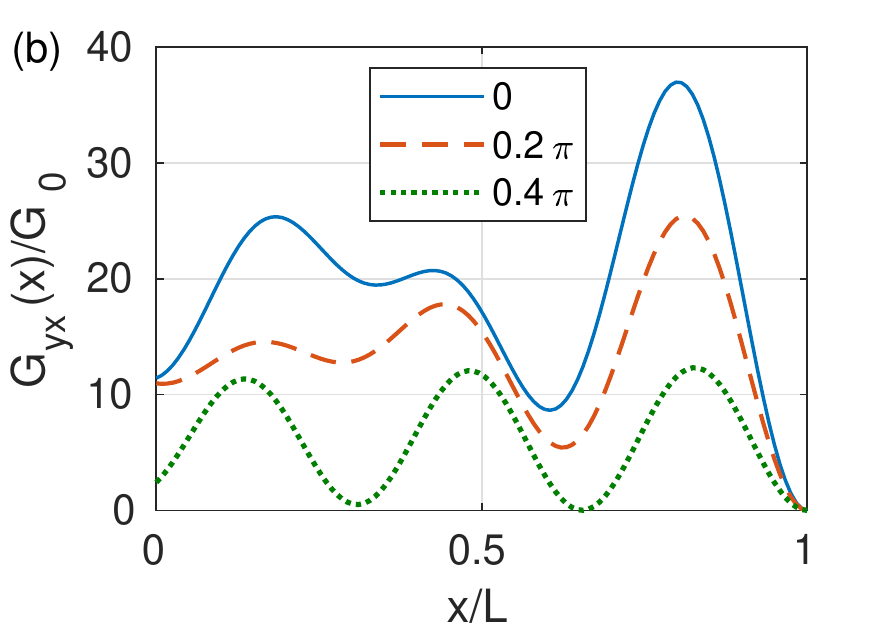}
 \caption{Zero bias transverse conductance as a function of the location for different choices $\phi$ indicated in the legend for (a) $\mu_t=\mu_b=0.5\mu_N$ and (b) $\mu_t=-\mu_b=0.5\mu_N$. Other parameters: $L=20\hbar v_F/\mu_N$, $b=0.2\mu_N$ and $m=0.025\mu_N/v_F^2$. }~\label{fig:gyxvsxneq}
\end{figure}

In Fig.~\ref{fig:gyxvsxneq}, we plot $G_{yx}(x)$ versus $x$ for a longer TI region with $L=20\hbar v_F/\mu_N$ for (a)~$\mu_t=\mu_b=0.5\mu_N$ and (b)~$\mu_t=-\mu_b=0.5\mu_N$ for different choices of $\phi$ to show the dependence of the transverse conductance on spatial location. Compared to Fig.~\ref{fig:gyxvsx}, the transverse conductance oscillates more as a function of $x$ in the range $0\le x\le L$ due to Fabry-P\'erot interference of the modes in TI. The relatively higher magnitude of transverse conductance in Fig.~\ref{fig:gyxvsxneq}(a) in comparison with that in Fig.\ref{fig:gyxvsx} is because of a higher value of  $\mu_t=\mu_b$. Further, we find that $G_{yx}(x)=G_{yx}(L-x)$ in the range $0\le x\le L$ when $\mu_t=\mu_b$, whereas $G_{yx}(L^-)=0$ always. 

Let us reason out analytically why $G_{yx}(L^-)=0$. The wavefunctions on the two TI surfaces and on the NM side at the location $x=L$ are related by the boundary condition eq.~\eqref{eq:bc} simplified to $\psi_t+\psi_b=\psi_N/c$ and $\psi_t-\psi_b=-dk_x\si_y\psi_N$, where $d=\hbar c/(mv_F)$. From these two equations, $\psi_N$ can be eliminated resulting in a relation between $\psi_t$ and $\psi_b$ from which it can be shown that $\psi_t^{\dagger}\si_x\psi_t=\psi_b^{\dagger}\si_x\psi_b$. This means the net transverse current $I_{\si,y}=ev_F(\psi_t^{\dagger}\si_x\psi_t-\psi_b^{\dagger}\si_x\psi_b)$ due to the two surfaces at $x=L^-$ is zero. 

\begin{figure}
 \includegraphics[width=6cm]{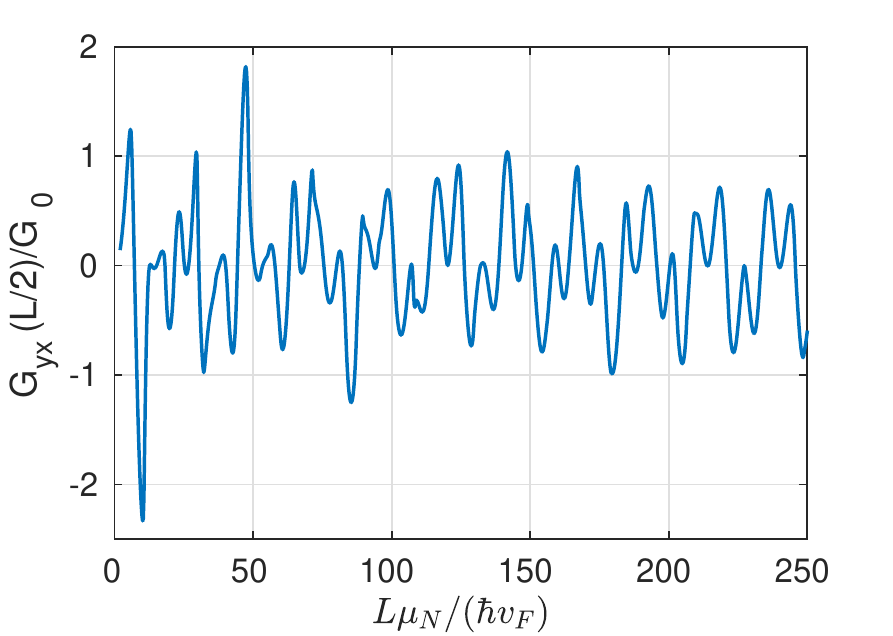}
 \caption{Zero bias transverse conductance evaluated at $x=L/2$ as a function of length $L$ for $\phi=0.2\pi$, $b=0.2\mu_N$, $\mu_t=\mu_b=0.5\mu_N$ and $m=0.25\mu_N/v_F^2$.}~\label{fig:GyxvsL}
\end{figure}

As a function of length $L$, the value of the transverse current $I_{\si,y}$ at given values of energy, angle of incidence $\th$ and spatial location (for instance at $x=L/2$) oscillates periodically due to Fabry-P\'erot type interference. But the transverse conductance which is obtained by integrating $I_{\si,y}$ over $\th$ need not be periodic in $L$ since the periods for different $\th$ will be different. However, $G_{yx}$  evaluated at $x=L/2$ oscillates about $0$ as a function of length $L$ as can be seen in Fig.~\ref{fig:GyxvsL}.

\subsection{Finite $L_y$}
\begin{figure}
 \includegraphics[width=4.2cm]{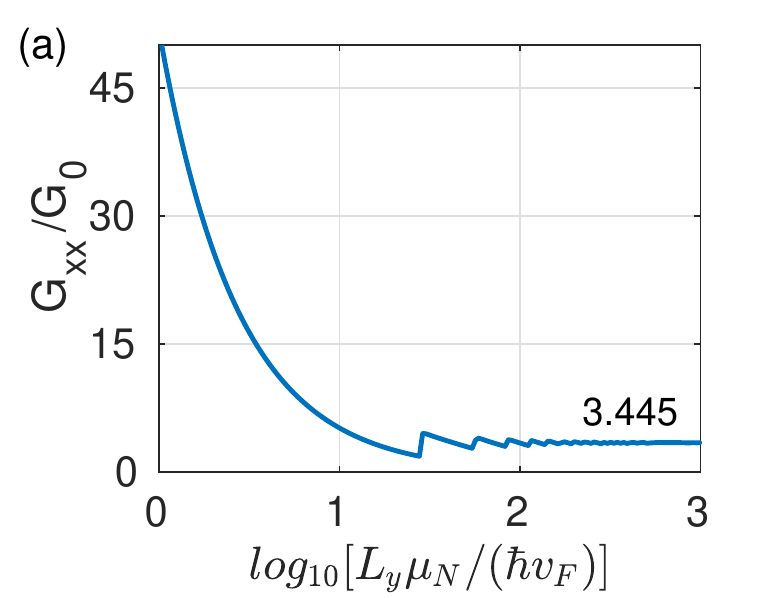}
 \includegraphics[width=4.2cm]{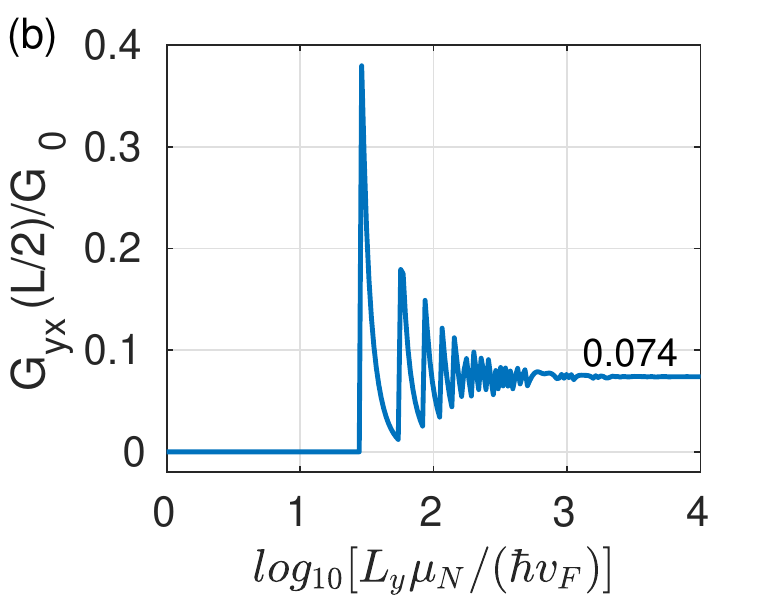}
 \caption{Dependence of zero bias- (a)~longitudinal and (b)~transverse conductances on $L_y$ for the choice of parameters: $L=5\hbar v_F/\mu_N$,  $\mu_t=\mu_b=0.1\mu_N$, $b=0.2\mu_N$, $m=0.025\mu_N/v_F^2$, $\phi=0.25\pi$. The saturation values of the conductances are mentioned in the figure. }\label{fig:G-Ly-eq}
\end{figure}

\begin{figure}
  \includegraphics[width=4.2cm]{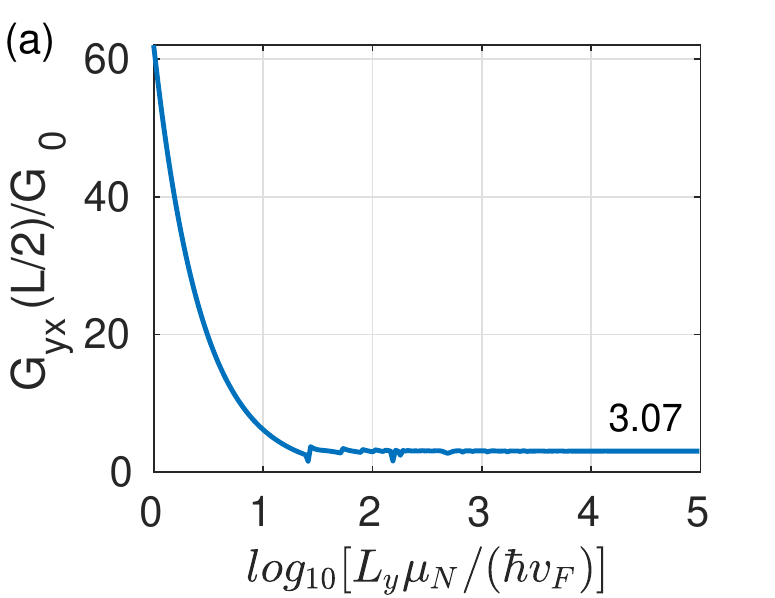}
  \includegraphics[width=4.2cm]{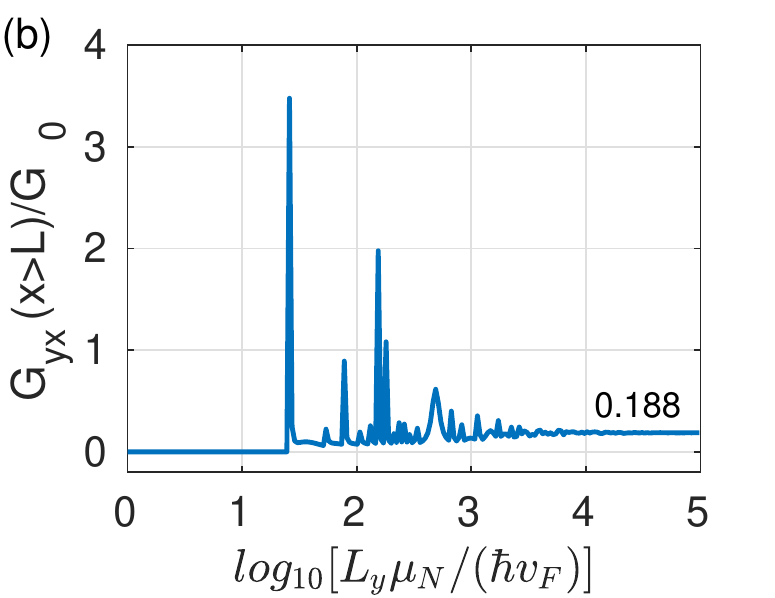}
  \caption{ The dependence of transverse conductances at (a) $x=L/2$ and (b) $x>L$ on $L_y$ for the choice of parameters: $L=5\hbar v_F/\mu_N$,  $\mu_t=-\mu_b=0.1\mu_N$, $b=0.2\mu_N$, $m=0.025\mu_N/v_F^2$, $\phi=0.25\pi$ and $E=0.2\mu_N$. The saturation values of the conductances are shown in  the figure.}\label{fig:G-Ly-neq}
 \end{figure}

Now, we turn to the case of finite $L_y$. The longitudinal and the transverse conductances for this case are given by eq.~\eqref{eq:G-finite}. From these formulae, it can be seen that as $L_y$ increases, the conductances draw contributions from more number of channels and  hence at large $L_y$ the conductances are proportional to $L_y$. Hence we plot the conductances in units of $G_0$ which is proportional to $L_y$. We first choose the parameters: $L=5\hbar v_F/\mu_N$,  $\mu_t=\mu_b=0.1\mu_N$, $b=0.2\mu_N$, $m=0.025\mu_N/v_F^2$, $\phi=0.25\pi$ and study the 
 dependence of $G_{xx}/G_0$ and $G_{yx}(x=L/2)/G_0$ as functions of $L_y$ in Fig.~\ref{fig:G-Ly-eq}. For $\mu_t=\mu_b$, the transverse conductance $G_{yx}(x>L)$ in the region $x>L$  is zero.  We can see that for large lengths, the two conductances saturate to the respective values in the limit of $L_y\to\infty$ that can be read from Fig.~\ref{fig:gvsVphib}~(c,d). For $\log_{10}[L_y\mu_N/(\hbar v_F)]<1.461$, there is only one channel participating in the transport and $G_{xx}/G_0$ increases as $L_y$ decreases since $G_0\propto L_y$. For the case of single channel, the contribution to transverse current  from the top and the bottom surfaces in the region $0<x<L$ are equal and opposite when $\mu_t=\mu_b$ and hence $G_{yx}(x)$ in this region is zero. 
Now, we turn to the case $\mu_t\neq\mu_b$. The longitudinal conductance shows features similar to the case $\mu_t=\mu_b$. But, the transverse conductance in the region $x>L$ is nonzero typically. For the choice of parameters:  $L=5\hbar v_F/\mu_N$,  $\mu_t=-\mu_b=0.1\mu_N$, $b=0.2\mu_N$, $m=0.025\mu_N/v_F^2$, $\phi=0.25\pi$, $E=0.2\mu_N$, we plot the  transverse conductances at $x=L/2$ and $x>L$ as functions of $L_y$ in Fig.~\ref{fig:G-Ly-neq}. A contrasting feature in this case compared to the case of $\mu_t=\mu_b$ is that the transverse conductance at $x=L/2$ for the values of $L_y$ corresponding to single channel is non-zero here. This can be understood by the following argument.   If $k_{xt},~k_{xb}$ are the $x$-components of wavenumbers on top and bottom TI surfaces, their Hamiltonians for single channel case ($k_y=0$) are $[(-\hbar v_Fk_{xt}+b_y)\si_y+b_x\si_x-\mu_t\si_0]$ and $[(\hbar v_Fk_{xb}+b_y)\si_y+b_x\si_x-\mu_b\si_0]$ respectively. It can be seen from here that when $\mu_t=\mu_b$, $\psi_t^{\dagger}\si_x\psi_t=\psi_b^{\dagger}\si_x\psi_b$ implying the transverse current $I_{\si,y}=ev_F(\psi_t^{\dagger}\si_x\psi_t-\psi_b^{\dagger}\si_x\psi_b)=0$. However, when $\mu_t\neq \mu_b$, the expectation values of $\si_x$ on the top and the bottom surfaces are not equal due to the terms $-\mu_{t/b}\si_0$ in the two Hamiltonians leading to a nonzero value of $I_{\si,y}=ev_F(\psi_t^{\dagger}\si_x\psi_t-\psi_b^{\dagger}\si_x\psi_b)$. Another feature of unequal chemical potentials on the two TI surfaces is that $G_{yx}(x>L)$ is generically nonzero. But for value of $L_y$ corresponding to the single channel case, $G_{yx}(x>L)=0$ since $G_{yx}(x>L)\propto k_y$ from eq.~\eqref{eq:G-finite} and $k_y=0$ for the only channel. 
The typical dependences of the conductances on $\phi$ for finite $L_y$ is similar to those observed for the case $\lim L_y\to\infty$, except for a difference in exact numerical values. 

\section{Discussion and Conclusion}~\label{sec:conclude}
We have essentially studied the phenomenon of PHE in TIs with the scattering theory approach when TI is connected to NM leads on either sides. We use the boundary condition for the NM-TI junction obtained by demanding the current conservation. The longitudinal and the transverse conductances are $\pi$-periodic in $\phi$ when $\mu_t=\mu_b$.  For angles $\phi$ close to $0$ or $\pi$, the longitudinal conductance decays with magnetic field whereas for angles $\phi$ close to $\pm\pi/2$, the longitudinal conductance decays with the magnetic field much slowly showing a slight periodic behavior with magnetic field strength at $\phi=\pm\pi/2$. Magnitude of the transverse conductance first increases with magnetic field, peaks and then decreases for angles $\phi$ close to but not equal to $0$ or $\pi$ whereas oscillates after an initial monotonic increase for angles close to $\pm\pi/2$. Such oscillations are rooted in Fabry-P\'erot type interference of the modes in the TI region between the two NM-TI junctions.  The transverse conductance depends on the spatial location and is zero in the right NM lead when $\mu_t=\mu_b$. When $\mu_t=-\mu_b$, the transverse conductance is nonzero though small in magnitude in the right NM region. We also find that when the width of the system $L_y$ is finite the scattering problem reduces to a one dimensional problem separated into a finite number of channels and  the conductances depend on the width of the system. The transverse conductance in the limit of small $L_y$ corresponding to a single channel, is zero at $x>L$ always whereas is zero in the region $0<x<L$ when $\mu_t=\mu_b$. 
The differential gating of the top and the bottom surfaces of the TI can be experimentally achieved which means $\mu_t$ and $\mu_b$ can be separately controlled~\cite{taskin17}. While many features in our results qualitatively agree with the experimental findings of Taskin et~al.~\cite{taskin17}, the angular dependence of the transverse resistance for the case of differentially gated top and bottom surfaces of the TI, the dependence of the conductances on the magnetic field strength and the dependence on width of the sample $L_y$ need to be probed experimentally.

\acknowledgements
Authors thank Bertrand Halperin, Karthik V. Raman and Archit Bhardwaj for useful discussions. AS thanks DST-INSPIRE Faculty Award (Faculty Reg. No.~:~IFA17-PH190) for financial support. 

\bibliography{ref_ti}

\end{document}